\documentclass[sigplan,nonacm]{acmart}

\usepackage[utf8]{inputenc}
\usepackage{algorithmic}
\usepackage{graphicx}
\usepackage{textcomp}
\usepackage{xcolor}
\usepackage{soul}
\usepackage{algorithm}
\usepackage{fancyvrb}
\usepackage{pifont}
\usepackage{subcaption}
\VerbatimFootnotes
\usepackage{booktabs}
\usepackage{url}
\newcommand{\RR}{\mathbb{R}}
\usepackage{microtype}
\usepackage{todonotes}

\usepackage{enumitem}
\setlist{nosep}

\usepackage{amsmath}

\usepackage{diagbox}

\usepackage{amsfonts,amssymb}

\begin{abstract}
Benefiting from the advancement of hardware accelerators such as GPUs, deep neural networks and scientific computing applications can achieve superior performance. Recently, the computing capacity of emerging hardware accelerators has increased rapidly, while memory bandwidth has not kept pace with this growth. This disparity exacerbates the gap between computing and memory, leading to inefficiencies on conventional algorithms, as they're likely to be converted from compute-bound to memory-bound. Symmetric eigenvalue decomposition (EVD), a critical operation in various research domains including scientific computing, deep learning training, and inference algorithms, exhibits suboptimal performance due to achieving less than 3\% hardware computing utilization on the H100 GPU. In this paper, we analyze the features of emerging hardware accelerators to identify the bottlenecks inherent in conventional EVD algorithms. To improve EVD performance, we propose several algorithmic optimizations aimed at solving the memory-bound problem and providing a better utilization of the rich computing capacity and parallelism on the emerging hardware accelerators. Experimentally, our proposed method demonstrates significant speedups on tridiagonalization, which is the main workload that takes over 90\% elapsed time of EVD, compared to the SOTA cuSOLVER tridiagonalization, achieving up to 10.1x, 7.5x, and 2.3x improvements on H100, A100, and RTX 4090 GPUs, respectively. And the end-to-end the performance of EVD solver is also up to 4.1x faster than cuSOVLER.
\end{abstract}

\begin{document}

% \title{Improving the Eigenvalue Decomposition Performance On New Generations of GPU Architectures}
\title{Extracting the Potential of Emerging Hardware Accelerators for Symmetric Eigenvalue Decomposition}
\author{Hansheng Wang, Lu Shi}
\affiliation{School of Computer Science and Engineering\\University of Electronic Science and Technology of China}
\email{{wanghansheng, 202411081638}@std.uestc.edu.cn}
\author{Zhekai Duan}
\affiliation{Department of Computer Science\\University of Edinburgh}
\email{s2085313@ed.ac.uk}
\author{Panruo Wu}
\affiliation{Department of Computer Science\\University of Houston}
\email{pwu7@uh.edu}
\author{Liwei Guo, Shaoshuai Zhang}
\affiliation{School of Computer Science and Engineering\\University of Electronic Science and Technology of China}
\email{{lwg, szhang94}@uestc.edu.cn}
\date{July 2019}

\maketitle
\pagestyle{plain}
\section{Introduction}
Driven by the need to train deep neural networks, particularly large language models containing billions of parameters, hardware accelerators such as GPUs and NPUs have evolved rapidly. For instance, compared to Nvidia's P100 GPU released in 2016, the H100 GPU achieves over 50x and 14x speedup in FP16 and FP64 precision, respectively. Leveraging the immense parallelism and computational capacity of these advanced hardware accelerators, various applications, including scientific computing~\cite{MAGMA}, can harness the power of GPUs to achieve performance gains that are orders of magnitude greater than those attainable with the latest generations of CPUs~\cite{LAPACK}.

However, the increase in memory bandwidth has not kept pace with the surge in computational capacity. For example, the P100 GPU offers 732 GB/s of memory bandwidth, while the H100 GPU provides 3430 GB/s. Despite the computational capacity growing by a factor of over 14x, the memory bandwidth has increased by less than 5x. This disparity presents significant challenges in designing hardware-accelerated algorithms, as some conventional algorithms may shift from being compute-bound to memory-bound (Figure~\ref{fig:roofline} illustrates this issue). In this paper, we use symmetric eigenvalue decomposition (EVD), a crucial problem in scientific computing and machine learning, to thoroughly explore these challenges.

Symmetric EVD is a fundamental matrix computation in numerical linear algebra with applications spanning quantum chemistry~\cite{QuantumChemistry}, quantum mechanics and physics~\cite{grimes1987solution, probert2011electronic}, and numerous machine learning and signal processing tasks~\cite{xsvm, tensorsvm, shampoo}. Computing EVD typically involves two steps: 1) tridiagonalization, which converts a given symmetric matrix $A$ to a tridiagonal matrix $T$ ($O(n^3) $ complexity); and 2) an iterative method such as the QR algorithm~\cite{QRAlgorithm} to compute the eigenvalues ($O(n^2)$ complexity). However, the state-of-the-art cuSOLVER~\footnote{\url{https://docs.nvidia.com/cuda/cusolver/}} implementation of tridiagonalization (\verb|cuSolverDnSytrd|) on GPUs achieves less than 3\% of the peak performance on the H100 GPU (1.8 TFLOPs out of 67 TFLOPs). Even the hybrid MAGMA~\cite{MAGMA} implementation, which employs a 2-stage tridiagonalization process (band reduction and bulge chasing)~\cite{2stage1}, fails to utilize more than 7\% of the computing capacity.

These observations prompt two critical questions: What are the reasons behind this suboptimal performance, and how to improve it. In this paper, we aim to addressing the two questions by proposing an optimized EVD solver designed for emerging hardware accelerators. Our experimental results demonstrate significant performance improvements over existing solutions, providing valuable insights into the efficient utilization of emerging hardware accelerators for matrix computations.

We consider our contributions of this paper to be:
\begin{itemize}
    % \item We benchmark and analyze the EVD solver on the H100 GPU using cuSOVLER and MAGMA, identifying the bottlenecks of the current SOTA EVD solvers.
    \item We analyze the features and evaluate the performance behaviours of the emerging hardware accelerators, and we use the SOTA EVD solver as an example to reveal the bottlenecks and difficulties of designing algorithms on these hardware. 
    \item We identify that, on emerging hardware accelerators, the SOTA band reduction process in tridiagonalization is memory-bound. To solve this problem, we propose a new band reduction algorithm that can significantly improve the inefficiency.
    % \item We propose a novel band reduction algorithm and a BLAS3 operation that leverage more computing resources on hardware accelerators compared to conventional band reduction algorithms.
    % \item We fully utilize the parallelism of hardware accelerators to perform bulge chasing process, which is normally considered not to benefit from the hardware accelerators, and get up to 7.9x speedup.
    \item We demonstrate the consensus, that the bulge chasing process in tridiagonalization is hard to be accelerated by hardware accelerators, is not correct. And we leverage the parallelism on hardware accelerators to obtain 7.9x speedup compared to conventional CPU-based implementations. 
    \item We implement and optimize the tridiagonalization which is the main workload of EVD solver, the experiments address that, compared to cuSOVLER and hybrid CPU-GPU MAGMA implementation, the proposed tridiagonalization is up to 10.1x, 6.0x speedup and the end-to-end EVD speedup is 4.1x and 3.7x, respectively.
    
\end{itemize}
The rest of the paper is organized as follows:  Section 2 explains the basic concepts of direct and two-stage tridiagonalization, as well as the bulge chasing process. Section 3 analyzes the hardware features, algorithmic bottlenecks, and addresses the motivations. Section 4 illustrates our proposed methods. Section 5 presents the implementation details and section 6 evaluates our implementations. Section 7 introduces related work regarding EVD on modern computer architectures. Finally, Section 8 draws conclusions and discusses future work.

\section{Background}
% \lwg{my greatest concern is the reviewers would not be able to comprehend the algorithm and such ...}
% \lwg{maybe we can merge the related work of tridiagnoalization here. Give a very brief and intuitive example on how it can be used.. Then bring in the equations ..}
To better illustrate our proposed method, we will simplify the background knowledge of the direct and 2-stage tridiagonalization process. This will provide a clearer understanding of the fundamental concepts and the improvements our method introduces.

\subsection{Tridiagonalization}
The tridiagonalization process is usually the pre-step to eigenvalue decomposition. The goal of tridiagonalization can be expressed as follows:
$$
A=Q^{-1}\times T \times Q,
$$
where $Q$ is an orthogonal matrix, and $T$ is a tridiagonal matrix. The conventional tridiagonalization~\cite{blockReduction} uses Householder reflection to eliminates the elements except the diagonal and the subdiagonal elements in one column. Unfortunately, on the modern computer architectures, the tridiagonalization is not efficient at all, because it contains too many BLAS2 operations (such as symmetric matrix vector multiplication (\verb|symv|)), which are not efficient and takes over 90\% of the total elapsed time. To solve this issue, 2-stage tridiagonalization is proposed and it's has a better utilization of hardware resources.

\subsection{2-stage Tridiagonalization}
\begin{figure}
    \centering
    \includegraphics[width=1.0\columnwidth]{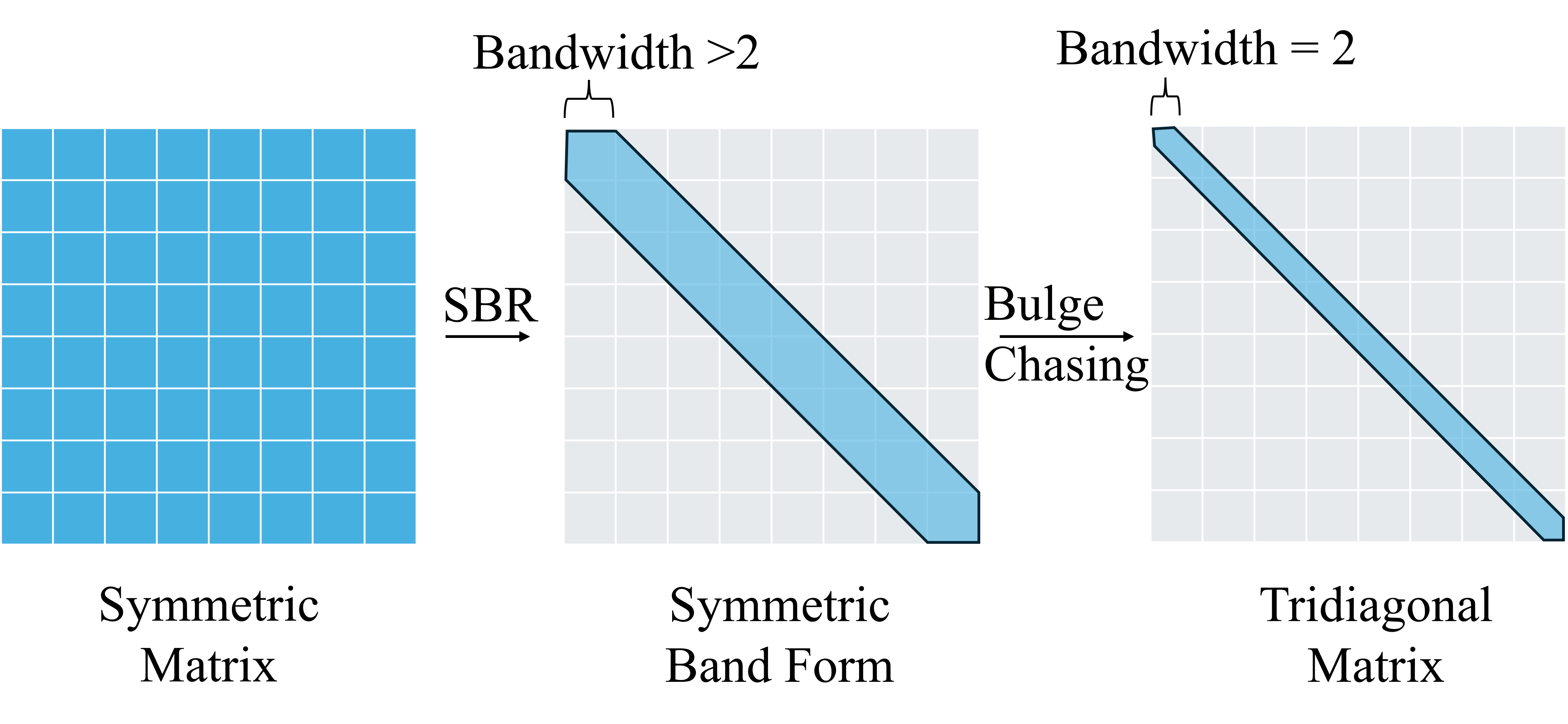}
    \caption{The 2-stage tridiagonalization process\label{fig:2-stage}}
\end{figure}

The 2-stage tridiagonalization has two successive processes, the first is called successive band reduction (SBR) that reduces the matrix to a band form: $A=Q^{-1}\times B\times Q$, where $Q$ is an orthogonal matrix and $B$ is a band matrix with bandwidth $b$. The 2nd stage is bulge chasing, it converts the band form matrix $B$ from 1st stage to a tridiagonal matrix $T$, see Figure~\ref{fig:2-stage} for a intuitive illustration. The 2-stage tridiagonalization is mathematically equivalent to direct tridiagonalization, but it provides more BLAS3 operations in SBR. Although the bulge chasing has many BLAS2 operations, the time complexity is much lower than SBR, so it doesn't take too much time to be performed. Therefore, the 2-stage is proved to be more efficient on multi-core architectures~\cite{2stage1,2stagesvd}.

\subsubsection{Successive Band Reduction}

The SBR process iteratively performs QR factorization and trailing matrix updates using the Symmetric Rank-2k update (\verb|syr2k|) routine, as illustrated in the left figure of Figure~\ref{fig:sy2sb_process}. The QR factorization aims to find the orthogonal transformation, eliminating the off-band entries while forming the matrices $Z$ and $Y$ required by trailing matrix update. Subsequently, the rest of the matrix is updated from both sides with the $Z$ and $Y$ matrices using \verb|syr2k|: $A=A-ZY^T-YZ^T$.

Compared to direct tridiagonalization, SBR is more efficient because it converts the column-by-column Householder transformation into QR factorization, which involves more BLAS3 operations. Although an additional bulge chasing process is required, the overall tridiagonalization performance is still improved because the time complexity of bulge chasing is only $O(nb^2)$.

\subsubsection{Bulge Chasing}
\begin{figure}
    \centering
    \includegraphics[width=1.0\columnwidth]{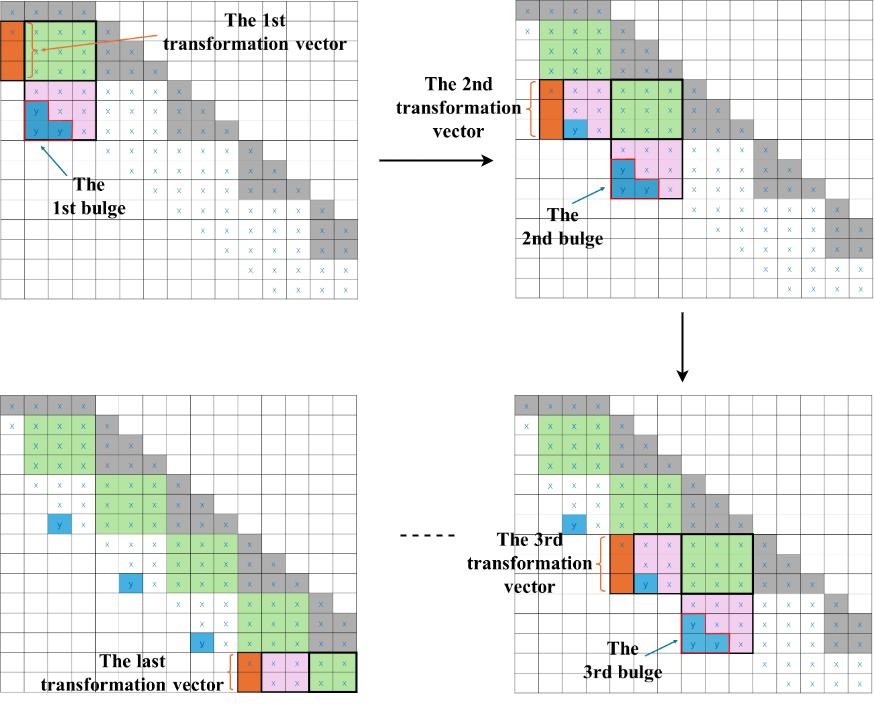}
    \caption{The 2-stage tridiagonalization process\label{fig:bcprocess}}
\end{figure}
The bulge chasing process reduces a band form matrix to a tridiagonal or bidiagonal matrix~\cite{BugleChasing}. In essence, the steps of bulge chasing are quite similar to tridiagonalization, as both iteratively apply Householder transformations to the trailing matrix. The key difference is that the input matrix for bulge chasing is a band matrix, allowing the computations to exploit the band structure and reduce the number of mathematical operations.

The steps of one sweep in bulge chasing are shown in Figure~\ref{fig:bcprocess}. It iteratively finds the Householder vectors and performs the Householder transformation to chase the bulge until the off-diagonal elements of one column are eliminated. In Figure~\ref{fig:bcprocess}, the orange columns denote the search for Householder vectors to eliminate elements in the current column. Once the Householder vector $v$ is formed, we use $H(v)=I-2vv^T/(v^Tv)$ to update the green blocks $G$ from the left and right sides that $G=H(v)^{-1}GH(v)$. The pink blocks $P$ can be updated from the left side only that $P=H(v)^{-1}P$, thereby creating a bulge denoted by the blue blocks. To chase down the bulge, we find the Householder vectors of the first column of the bulge and repeat the aforementioned steps until the bulge is swept down to the last column.

The above steps demonstrate one sweep of the bulge chasing process. In fact, to fully reduce the band form matrix to a tridiagonal matrix, $n-2$ sweeps if the given matrix's size is $n\times n$. Obviously, the bulge chasing remains memory-bound because $b$ is typically much smaller than $n$. Consequently, even in recent research from 2018, Gates et al.\cite{2stagesvd} assert that "as this stage (bulge chasing) has limited parallelism, is close to memory bandwidth limited, and is already optimized for the CPU caches, it would not benefit much, if any, from an accelerator-based implementation" (Section 4.2). Therefore, all bulge chasing implementations are deployed on CPUs, as seen in LAPACK\cite{LAPACK}, PLASMA~\cite{PLASMA}, and MAGMA~\cite{MAGMA}, named \verb|sb2st| routine.

\subsection{Iterative Methods}
After obtaining the tridiagonal matrix, iterative methods are used to get eigenvalues.
The most popular iterative methods for computing eigenvalues and eigenvectors are the QR algorithm~\cite{QRAlgorithm} and divide and conquer (D\&C)~\cite{DivideAndConquer}, both of which are included in numerical linear algebra libraries such as cuSOLVER. The primary difference between these methods is their computational complexity: if only eigenvalues are needed, the QR algorithm requires $O(n^2)$ operations,whereas D\&C requires $O(n^3)$ because it also generates eigenvectors during the iterations but it has better parallelism than QR algorithm.

% \subsubsection{QR Algorithm}

% Given a symmetric tridiagonal matrix $A\in \RR^{n*n}$, the QR algorithm iteratively performs the following steps:
% $$
% \text{QR factorization:} A_{(i)}\rightarrow Q\times R,
% $$
% $$
% \text{GEMM:}A_{(i+1)}\leftarrow R\times Q.
% $$
% And finally, the matrix $A$ can be reduced to a diagonal matrix and the diagonal elements are the eigenvalues.

% The above iterations can be simplified if eigenvectors are not needed, which is called implicit QR algorithm. Notice that $R$=$Q^TA_{(i)}$, then we have $A_{(i+1)}=RQ=Q^TA_{(i)}Q$, which means we don't have to form the upper triangular matrix $R$, orthogonal transformation (Givens rotation) is all we need. 

% However, these iterations can be very slow when two eigenvalues are not well-separated. To accelerate the convergence speed, shifting can be applied:
% $$
% A_{(i)}-\sigma I \rightarrow Q\times R,
% $$
% $$
% A_{(i+1)}\leftarrow R\times Q+\sigma I.
% $$
% The typical selection of $\sigma$ is using Wilkinson shift~\cite{WilkinsonShift}, which picks one eigenvalue of the bottom-right $2\times 2$ submatrix. The Wilkinson shift QR algorithm always converge and with at least quadratic rate.

% \subsubsection{Divide and Conq}

\section{Motivation}

In this section, we will show the motivations of this paper by addressing hardware features and study the bottlenecks. By understanding the features and bottlenecks, we can propose targeted optimizations to improve the efficiency of EVD solvers, especially the tridiagonalization process.

\subsection{The Features of Emerging Hardware Accelerators}
% \begin{table}[]

% \centering
% \begin{tabular}{ |c|c|c| } 
% \hline
% GPU & Peak FP64 Performance & Memory Bandwidth \\
%  \hline
%  P100  & 4.7 TFLOPs & 732 GB/s   \\
% \hline
%  V100  & 7.0 TFLOPs & 900 GB/s  \\
% \hline
%  A100 &  19.5 TFLOPs & 1935 GB/s  \\
% \hline
%  H100  & 67.0 TFLOPs & 3430 GB/s   \\
% \hline
% \end{tabular}
% \caption{The peak performance and memory bandwidth comparison between P100, V100, A100, and H100}
% \label{tbl:gemm_generation}
% \end{table}
% \lwg{this section can be moved to motivation sections ...}
The emerging hardware accelerators are designed to handle the computational intensity of DNNs, especially Transformer-based DNNs~\cite{Transformer}, which typically involve a substantial amount of matrix multiplications (GEMMs). For instance, Nvidia's latest GPUs feature Tensor Cores that excel in low or mixed-precision GEMMs, offering nearly 1 PFLOPs peak performance for FP16 precision GEMMs on H100 GPU. Additionally, high-precision (FP64) GEMMs can also be executed on Tensor Cores, delivering performance comparable to FP32 precision GEMMs.

While the computing capacity of hardware continues to increase rapidly, the memory bandwidth is also improving, but at a slower pace. This discrepancy leads to a widening memory-compute gap.
\begin{figure}
    \centering
    \includegraphics[width=1.0\columnwidth]{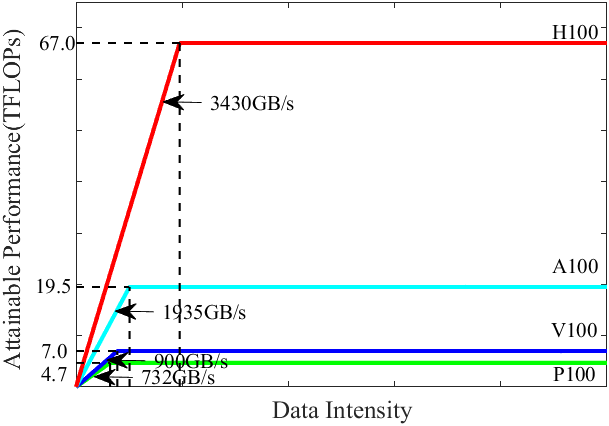}
    \caption{The FP64 precision roofline model of 4 GPU generations}
     \label{fig:roofline}
\end{figure}

To better understand the difficulties in fully utilizing the performance capabilities of emerging hardware, we employ the roofline model~\cite{williams2009roofline} and present Figure~\ref{fig:roofline} to illustrate the performance differences across various generations of Nvidia GPUs. As depicted, achieving peak performance on the latest GPUs, such as the H100 and A100, necessitates a higher data intensity. Having these features, however, conventional numerical linear algebra algorithms, like the EVD solver that utilize blocking algorithms, do not achieve the requisite data intensity to fully exploit the computational power of these advanced GPUs, which means compute-bound algorithms can be converted to memory-bound on new hardware. Thus, to understand the reasons behind the above conversion, we use EVD solver as an example and address the performance bottlenecks in the following section.

\subsection{Performance Bottleneck}
Extensive literature indicates that two-stage tridiagonalization outperforms direct tridiagonalization, even when forming eigenvectors~\cite{2stage1,2stage2,2stage3}. Thus, we will not delve deeply into the comparative analysis of these two approaches, but rather, we will focus on the bottlenecks of the state-of-the-art 2-stage tridiagonalization algorithm.
% It is noteworthy that for smaller matrix sizes (less than $8192 \times 8192$), cuSOLVER tridiagonalization (\verb|cuSolverDnDsytrd|) outperforms MAGMA~\cite{MAGMA} due to better utilization of CUDA cores and the elimination of data movement between the device and host.
% \lwg{donot have to mention cuSolver here? Otherwise reviewer may say: your approach is ad hoc ... some framework does not have such bottlenecks... }

% \subsubsection{Data Movement Bottleneck}
% The only GPU-based two-stage tridiagonalization algorithm is implemented in MAGMA~\cite{MAGMA}. However, only the \verb|syr2k| operations in SBR (MAGMA \verb|sy2sb| routine) are executed on GPUs, necessitating unavoidable data transfers between the host and device. While some of these data movements can overlap with computations, there remains a non-trivial overhead. A straightforward solution would be to perform all computations on the GPU to eliminate data movement. However, MAGMA relies on LAPACK's QR factorization routine, which is executed on the CPU. Additionally, prior research favors performing bulge chasing on the CPU rather than the GPU, leading to a hybrid implementation that further exacerbates the data movement bottleneck.

\subsubsection{Band Reduction Bottleneck}
As we mentioned before, the 1st stage of 2-stage tridiagonalization is SBR, which reduces a symmetric matrix to band form. This involves designating a bandwidth $b$, followed by QR factorization on a tall and skinny matrix of size $n \times b$, and \verb|syr2k| operations of size $n \times n \times b$. Despite increasing the proportion of BLAS3 operations compared to direct tridiagonalization, this method suffers from a critical performance limitation: the bandwidth $b$ is strictly equal to the block size $nb$. This necessitates a delicate balance between the first and second stages of tridiagonalization, as a larger $b$ results in slower bulge chasing, while a smaller $b$ leads to slower band reduction. Currently, MAGMA defaults to a bandwidth of 128 or 256 (although we find the $b=64$ can give the best performance on H100 and A100 GPU) to optimize performance; however, the \verb|syr2k| operation cannot achieve peak hardware performance due to the tall and skinny matrix shape. Connected to the aforementioned roofline model (Figure~\ref{fig:roofline}), current algorithmic design of SBR cannot provide enough data intensity and result in memory-bound implementation. And to fully utilize the hardware accelerators, it's necessary to figure out how to provide larger and more square \verb|syrk2k| operations. 

Quantitatively, Table~\ref{tbl:syr2k_perf} presents the state-of-the-art cuBLAS \verb|syr2k| performance (also used by MAGMA) for various $n$ and $k$ on an H100 GPU. The data reveals that only larger values of $k$ (corresponding to bandwidth $b$) deliver satisfactory performance. However, to balance the performance between SBR and bulge chasing, the bandwidth $b$ is typically set to 128 or 256, as larger $b$ significantly slows down bulge chasing. Consequently, the overall performance of SBR remains suboptimal and memory-bound, even with an increased proportion of BLAS3 operations.

\begin{table}
\centering
 \begin{tabular}[t]{ccccc} 
 \toprule
 \multicolumn{5}{c}{\includegraphics[height=2.0cm]{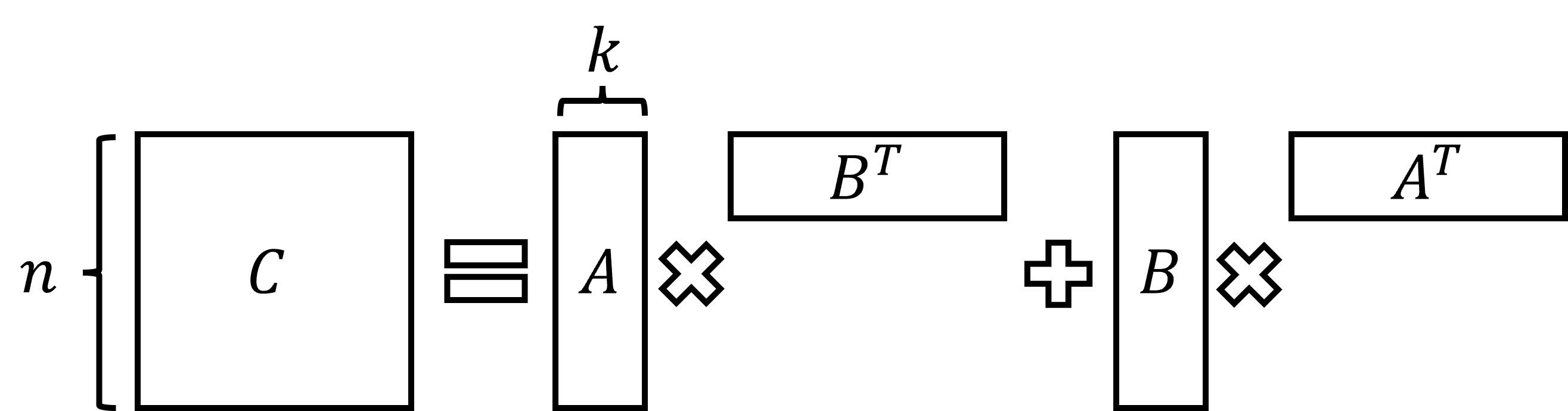}} \\
 \cmidrule{1-3} \cmidrule{4-5}
 $k$  & $n=4096$ & $n=8192$  & $n=16384$  & $n=32768$  \\
 \midrule
 16 & 0.09 &0.43  &1.60  &3.58   \\
 32 & 0.18 &0.86  &3.20  &7.02   \\
 
 64 & 0.38 &1.71  &6.19  &12.78   \\
 
 128 & 0.76 &3.39  &11.47  &21.05   \\
 
 256 & 1.42 &6.41  &18.83  &30.13  \\
 
 512 & 2.29 &11.57  &27.58 &38.31  \\

 1024 & 5.77 &18.91  &35.23  &42.86 \\
 
 2048 & 8.54 &27.21  &40.82  &45.36 \\
 
 4096 & 13.72 &34.59  &43.65  &45.54 \\
 
 \bottomrule
\end{tabular}
\caption{The performance of SYR2K on H100 GPU with different input sizes ($n$ and $k$) in TFLOPs}
\label{tbl:syr2k_perf}
\end{table}

\subsubsection{Bulge Chasing Bottleneck}
\begin{figure}
    \centering
    \includegraphics[width=1.0\columnwidth]{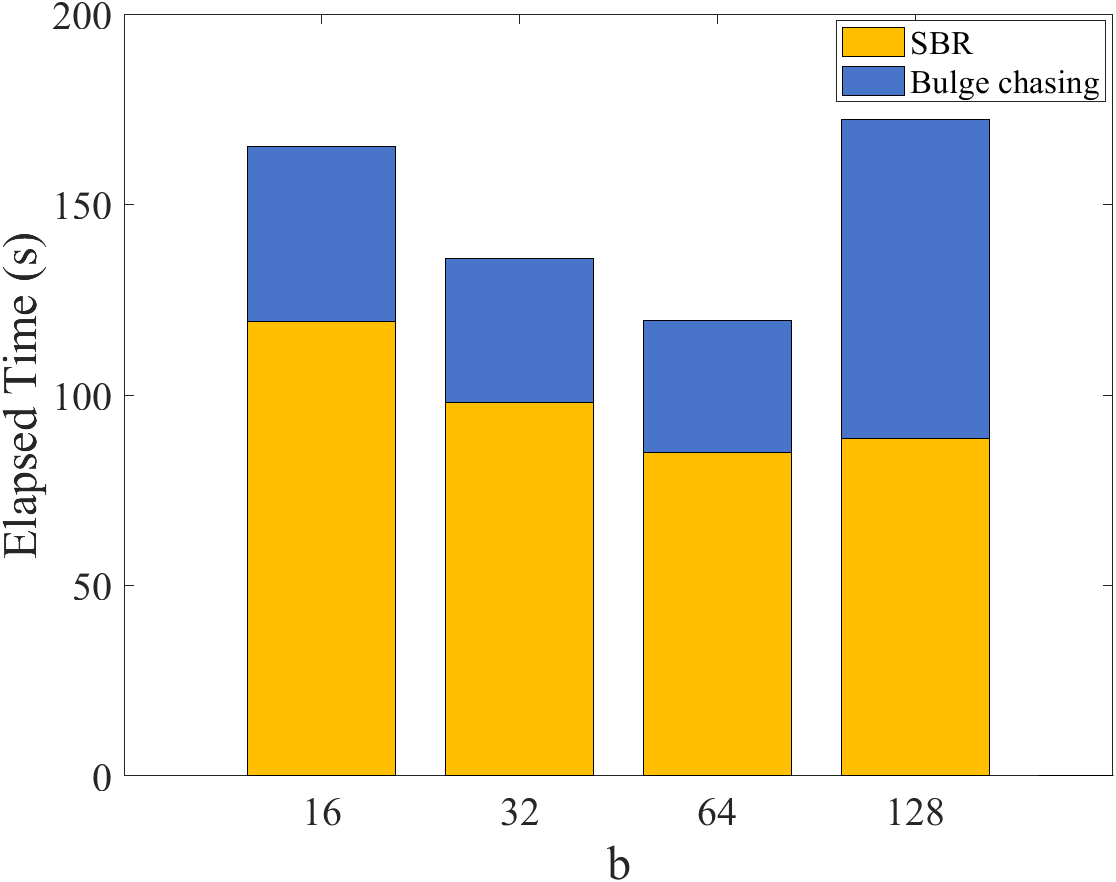}
    \caption{The MAGMA 2-stage tridiagonalization performance given a matrix $A\in \RR^{65536\times 65536}$ with different bandwidth $b$ on H100 GPU}
     \label{fig:magma_sy2tr}
\end{figure}

The bulge chasing process, which comprises numerous BLAS1/2 operations, is already memory-bound and difficult to be parallelized on old hardware accelerators. As a result, it cannot exhibit optimal performance on GPUs~\cite{2stage1,2stagesvd}. Our experimental evaluation (Figure~\ref{fig:magma_sy2tr}) demonstrates that for a matrix of size $65536 \times 65536$, the bulge chasing process, consuming over one-fourth of the tridiagonalization elapsed time, represents a significant performance bottleneck.

\subsection{Opportunities}
% \lwg{section name: Summary -> Opportunities}
In this section, we discuss the hardware features of the latest GPUs: 
the memory speed scales much slower than that of computations for evolving GPUs, which results in an enlarging gap between compute capacity and memory bandwidth.
This gap necessitates higher data intensity to achieve peak performance, complicating algorithm design. Based on our analysis of the performance bottlenecks, the conventional EVD algorithms, which is efficient on previous computer architectures, fails to handle the larger memory-compute gap and rich parallelism of emerging hardware, leading to a memory-bound implementations.

As a result, the features and bottlenecks leave us some opportunities to outperform the SOTA algorithms which are memory-bound on emerging hardware accelerators. In terms of EVD solver, in cuSOVLER and MAGMA implementations, given a matrix size $50K\times 50K$ on H100 GPU, the tridiagonalization process takes over 95\% (78.5 seconds out of 80.4 seconds), and 60\% (43.5 seconds out of 73.5 seconds), respectively. Thus, the EVD solver's performance will boost significantly if we can optimize the tridiagonalization quite well by utilizing the computing capacity and parallelism on new hardware. And we'll propose a new method of tridiagonalization to increase data intensity and handle the memory-bound problem in the following section.

% The conventional 2-stage tridiagonalization algorithm struggles to fully exploit the hardware's capabilities due to several performance bottlenecks. Completing the entire Eigenvalue Decomposition (EVD) requires iterative methods such as QR algorithm and divide and conquer, which, despite its $O(n^2)$ time complexity, has sufficient parallelism to be efficiently executed on GPUs, accounting for only around 3\% \todo{ref to fig} of the elapsed time—thus, it is not considered another bottleneck.

% In the following section, we present our proposed methods to increase data intensity, alongside implementations and optimizations designed to maximize hardware computational capacity.

\section{Methods}
% \lwg{Try not to mention MAGMA's implementations, which makes the work seem ad hoc}
Given the aforementioned bottlenecks, it is clear that although 2-stage tridiagonalization exhibits better data locality than direct tridiagonalization, the speedup is only around 1.6x (Figure~\ref{fig:tr_perf_h100}). The primary challenge lies in designing efficient SBR and bulge chasing on emerging hardware accelerators.

\subsection{Detached Band Reduction}
Each iteration in the conventional band reduction algorithm involves a tall and skinny QR (TSQR) factorization and trailing matrix updates using the \verb|syr2k| routine. There is extensive literature on performing fast TSQR on GPUs~\cite{QRTensorCore, CAQR, reconstructingHouseholderVectorsInCAQR}, which we can leverage directly. With the TSQR algorithm, the critical path is the trailing matrix update. As shown in Table~\ref{tbl:syr2k_perf}, typical bandwidth selections of 128 or 256 achieve less than 30\% of peak performance on the H100 GPU. Therefore, significantly improving \verb|syr2k|'s performance would yield substantial gains in band reduction.

One straightforward method is to increase the bandwidth $b$. From Table~\ref{tbl:syr2k_perf}, only a $k$ value of 1024 or larger in the \verb|syr2k| operation achieves satisfactory trailing matrix update performance. However, increasing the bandwidth indiscriminately degrades the bulge chasing performance, as seen in Figure~\ref{fig:magma_sy2tr} where even $b=128$ significantly increases the bulge chasing time cost.
\begin{figure}
    \centering
    \includegraphics[width=1.0\columnwidth]{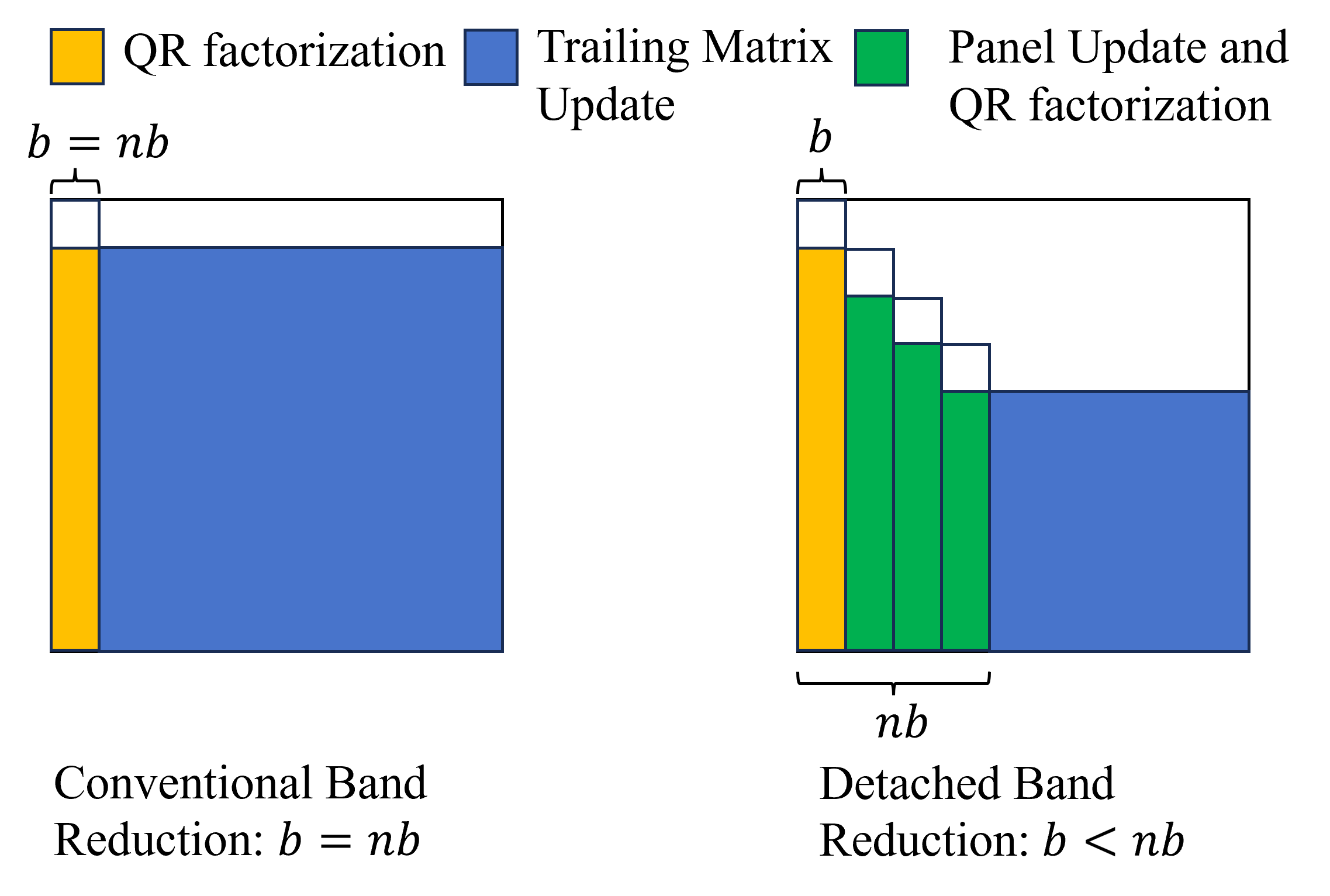}
    \caption{The differences between conventional and detached Band Reduction\label{fig:sy2sb_process}}
    
\end{figure}

To address this issue, we propose a new band reduction algorithm called Detached Band Reduction (DBR), and the key idea is to decouple the bandwidth $b$ from the blocksize $nb$ (Figure~\ref{fig:sy2sb_process}).
% \lwg{The key idea is to ...}
For a matrix $A \in \mathbb{R}^{n \times n}$, with bandwidth $b$ and block size $nb$, the algorithm is illustrated in Algorithm~\ref{alg:sbr}. If $b$ equals $nb$, DBR degrades to SBR.

\begin{algorithm}
\caption{Detached Band Reduction}\label{alg:sbr}
\begin{algorithmic}[1]
\REQUIRE A symmetric matrix $A\in \RR^{n\times n}$, bandwidth $b$ and blocksize $nb$, where $b\leq nb$
\ENSURE $A$ is reduced to a band form matrix with bandwidth $b$
\FOR{$i=1:nb:n$}
    \FOR{$j=i:b:nb$}
    \STATE $[W, Y, R] \leftarrow$ $QR(A_{panel})$
    \IF{$j+b<nb$}
         \STATE $Z \leftarrow AW-\frac{1}{2}YW^TAW$     
        \STATE $A\leftarrow A-ZY^T-YZ^T$
        \textcolor{red}{\%Only needed panel is udpated}
    \ENDIF
    \ENDFOR
    \STATE Accumulate matrix $Y$ and $Z$
    \STATE $A\leftarrow A-ZY^T-YZ^T$ \textcolor{red}{\%Trailing matrix update} 
\ENDFOR
\STATE Compute the orthogonal matrix $Q$ if needed
\end{algorithmic}
\end{algorithm}

DBR's primary advantage over SBR is the decoupling of blocksize $nb$ from bandwidth $b$, offering greater flexibility in selecting and adjusting these parameters. In SBR, the typical choice is $b=nb=64$, whereas DBR allows configurations like $b=32$ and $nb=1024$. This algorithmic optimization provides two significant performance benefits: 1) enabling much larger $k$ values in trailing matrix updates using the \verb|syr2k| routine; 2) allowing smaller bandwidth $b$ to reduce the time complexity of the subsequent bulge chasing process. These advantages lead to substantial performance improvements in band reduction, even with smaller bandwidths.

\subsection{Bulge Chasing Using Hardware Accelerators}
It is widely accepted that the two-stage tridiagonalization is faster on modern architectures because it converts many BLAS2 operations into BLAS3 operations. However, previous research asserts that the bulge chasing process, being limited in parallelism and close to memory bandwidth limits, would not benefit significantly from an accelerator-based implementation, as it is already optimized for CPU caches (Section 4.2 in~\cite{2stagesvd}). This explains why existing 2-stage tridiagonalization/bidiagonalization algorithms use CPUs for bulge chasing~\cite{2stage3, 2stagesvd}.

In this section, we refute these claims by demonstrating that the bulge chasing process, despite being memory-bound, still exhibits enough parallelism to benefit significantly from hardware accelerators. This parallelism is evident in two aspects: 1) inter-kernel parallelism (e.g., pipelining); 2) intra-kernel parallelism (e.g., Householder transformations).
% \lwg{Inter-kernel parallelism and intra-kernel parallelism.. The below says both are straightforward, which may get criticized...}

\begin{figure}
    \centering
    \includegraphics[width=1.0\columnwidth]{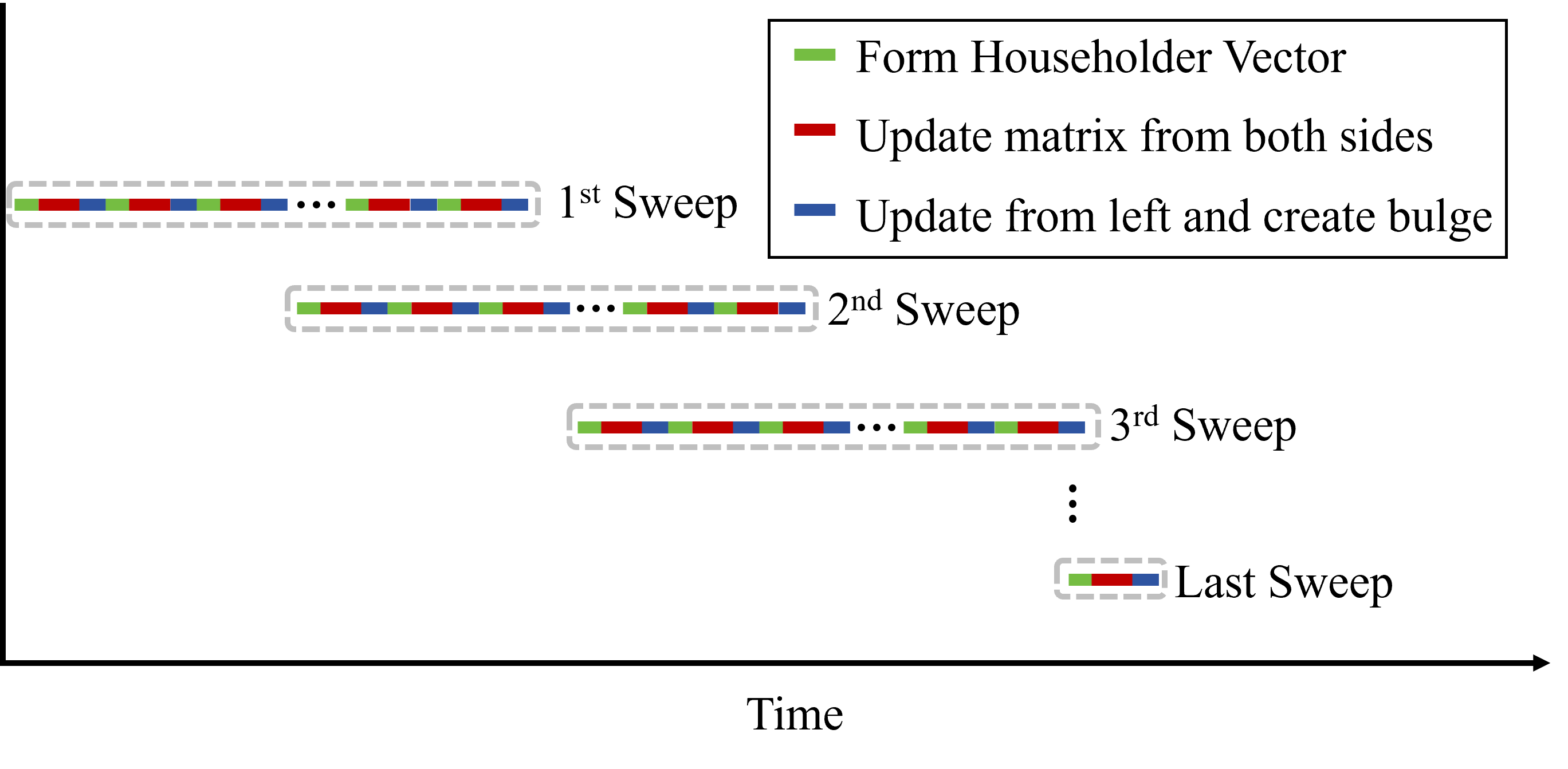}
    \caption{An illustration of the pipelining between successive sweeps in bulge chasing process\label{fig:bcpipeline}}
    
\end{figure}

The inter-kernel parallelism is enabled because different sweeps lack data dependencies, allowing for pipelining. To illustrate this pipeline, we categorize operations into three types: Householder vector generation, matrix update from both sides, and matrix update from the left side (creating a bulge), denoted by different colors in Figure~\ref{fig:bcpipeline}. We observe that when the $i$-th sweep completes three cycles, the $(i+1)$-th sweep can safely start without any data dependency. This allows one thread block to handle one sweep while other thread blocks wait and execute sequentially.

% \begin{algorithm}
% \caption{Parallel bulge chasing on GPU}\label{alg:bulge_chasing}
% \begin{algorithmic}[1]
% \REQUIRE A band form matrix $A\in \RR^{n\times n}$ with bandwidth $nb<<n$
% \ENSURE $A$ is reduced to a tridiagonal matrix
% \FOR{$i=1:1:n$}
%     \STATE $i+1$-th sweep waits for the $i$-th sweep's synchronization flag \textcolor{red}{\% The sweeps can be executed in parallel}
%     \STATE Compute Householder vectors $v_i$
%     \STATE $H(v_i)\leftarrow I-2\frac{v_iv_i^T}{v_i^Tv_i}$
%         \STATE $A_g\leftarrow=H(v_i)^{-1}A_gH(v_i)$ \textcolor{red}{\% Update matrix from both sides (green block in Figure~\ref{fig:bcprocess})}
%         \STATE $A_p\leftarrow=H(v_i)^{-1}A_p$ \textcolor{red}{\% Update matrix from left side (pink block in Figure~\ref{fig:bcprocess}) and create a bulge (blue block in Figure~\ref{fig:bcprocess})}
%     \FOR{$j=i+1:b:n$}
%         \STATE Compute Householder vectors $v_j$
%         \STATE $H(v_j)\leftarrow I-2\frac{v_jv_j^T}{v_j^Tv_j}$
%         \STATE $A_g\leftarrow=H(v_j)^{-1}A_gH(v_j)$ 
%         \STATE $A_p\leftarrow=H(v_j)^{-1}A_p$ 
%     \ENDFOR
% \ENDFOR
% \STATE Compute the orthogonal matrix $Q$ if needed
% \end{algorithmic}
% \end{algorithm}

\begin{algorithm}
\caption{Parallel bulge chasing on GPU}\label{alg:bulge_chasing}
\begin{algorithmic}[1]
\REQUIRE A band form matrix $A\in \RR^{n\times n}$ with bandwidth $nb<<n$
\ENSURE $A$ is reduced to a tridiagonal matrix
\STATE volatile int $gCom[n]={0}$;
\STATE \textcolor{red}{\% The sweeps can be executed in parallel}
\FOR{$i=1:1:n$}
    \STATE opCol = $i$  \textcolor{red}{\% $i$ represents begin column index of $i$-th sweep}
    \WHILE{$opCol<n$}
        \WHILE{(1 != i) \&\& ($opCol+2*b> gCom[i-1]$)}
        \STATE    continue
        \ENDWHILE

        \STATE Compute Householder vectors $v_i$
        \STATE $H(v_i)\leftarrow I-2\frac{v_iv_i^T}{v_i^Tv_i}$
            \STATE $A_g\leftarrow=H(v_i)^{-1}A_gH(v_i)$ \textcolor{red}{\% Update matrix from both sides (green block in Figure~\ref{fig:bcprocess})}
            \STATE $A_p\leftarrow=H(v_i)^{-1}A_p$ \textcolor{red}{\% Update matrix from left side (pink block in Figure~\ref{fig:bcprocess}) and create a bulge (blue block in Figure~\ref{fig:bcprocess})}
        \FOR{$j=i+1:b:n$}
            \STATE Compute Householder vectors $v_j$
            \STATE $H(v_j)\leftarrow I-2\frac{v_jv_j^T}{v_j^Tv_j}$
            \STATE $A_g\leftarrow=H(v_j)^{-1}A_gH(v_j)$ 
            \STATE $A_p\leftarrow=H(v_j)^{-1}A_p$ 
        \ENDFOR
        
        \STATE $opCol=opCol+b$ \textcolor{red}{\% update process position of $i$-th sweep}
        \STATE $gCom[i]=opCol$ \textcolor{red}{\% update sync variable}
    \ENDWHILE
\ENDFOR
\STATE Compute the orthogonal matrix $Q$ if needed
\end{algorithmic}
\end{algorithm}

The intra-kernel parallelism is primarily leveraged in Householder vector generation and Householder transformations. By launching multiple threads, the transformation can be computed in parallel to achieve satisfactory performance. See Algorithm~\ref{alg:bulge_chasing} for details.

Despite being memory-bound rather than compute-bound, bulge chasing can significantly benefit from emerging hardware accelerators, such as H100, A100, and RTX 4090 GPUs, by fully exploiting the parallelism between different sweeps and within CUDA kernels. Compared to the CPU implementation, the bulge chasing using hardware accelerators is nearly 8.0x faster than the CPU-based implementation (Figure~\ref{fig:bc_perf_comp}).

\section{Implementation and Optimization}

While our proposed methods can accelerate the EVD on GPUs, effective implementation and technical optimization are crucial to avoid wasting GPU resources. In this section, we discuss our kernel-level design strategy, including optimizations for DBR, BLAS3 routines, and bulge chasing.

\subsection{DBR Implementation}

As shown in Algorithm~\ref{alg:sbr}, the first step is panel QR factorization. Extensive research exists on implementing TSQR on GPUs, so we will not delve into it here; see~\cite{QRTensorCore, reconstructingHouseholderVectorsInCAQR} for high-performance TSQR implementations and the reconstruction of the WY representation.

The subsequent step is panel updates between $b$ and $nb$  (Line 6 in Algorithm~\ref{alg:sbr}). Similar to \verb|syr2k| performance (Table~\ref{tbl:syr2k_perf}), The GEMMs shapes and sizes also affect the performance.the shapes and sizes of GEMMs significantly affect performance. Unfortunately, the GEMMs in panel updates typically have a dimension equal to $b$, resulting in suboptimal performance. To address this, we can use a recursive strategy to enlarge these dimensions.

For illustration, consider $b=32$ and $nb=256$ in DBR. Non-recursive panel updates generate 7 GEMMs with $k=32$, while recursive panel updates produce 4 GEMMs with $k=32$, 2 GEMMs with $k=64$, and 1 GEMM with $k=128$. This modification does not increase the total computations but converts GEMMs into more square shapes, increasing FLOPs efficiency.

\subsection{Symmetric Rank-2k Update Optimization}

Although algorithmic design can speed up trailing matrix updates using cuBLAS \verb|Dsyr2k| routines, our experimental results show that cuBLAS routines still consume excessive time (over 90\% of DBR time). Therefore, optimizing \verb|syr2k| is crucial.

Testing cuBLAS \verb|Dsyr2k| on an H100 GPU reveals it achieves less than 60\% (36 TFLOPs) of the H100 peak FP64 precision. Additionally, a severe bug reduces \verb|Dsyr2k| performance to 4 TFLOPs when $n \geq 49152$.

To improve performance, we propose a recursive formulation that maximizes the size of internal GEMMs:
\begin{equation}
\label{eqa:syr2k}
\left[ \begin{array}{c|c}
C_{11} & C_{12} \\ \hline
C_{21}  & C_{22}
\end{array}   \right] = 
\left[ \begin{array}{c}
A_{11} \\ \hline
A_{21}
\end{array}   \right]
\left[ \begin{array}{c|c}
B_{11}^T & B_{21}^T
\end{array}   \right]+\left[ \begin{array}{c}
B_{11} \\ \hline
B_{21}
\end{array}   \right]
\left[ \begin{array}{c|c}
A_{11}^T & A_{21}^T
\end{array}   \right]
\end{equation}
This results in: 
$$
SYR2K: C_{11}=A_{11}B_{11}^T+B_{11}A_{11}^T
$$
$$
GEMM: C_{21}=C_{12}^T=A_{21}B_{11}^T+B_{21}A_{11}^T
$$
$$
SYR2K: C_{22}=A_{21}B_{21}^T+B_{21}A_{21}^T
$$

\begin{figure}
    \centering
    \includegraphics[width=1.0\columnwidth]{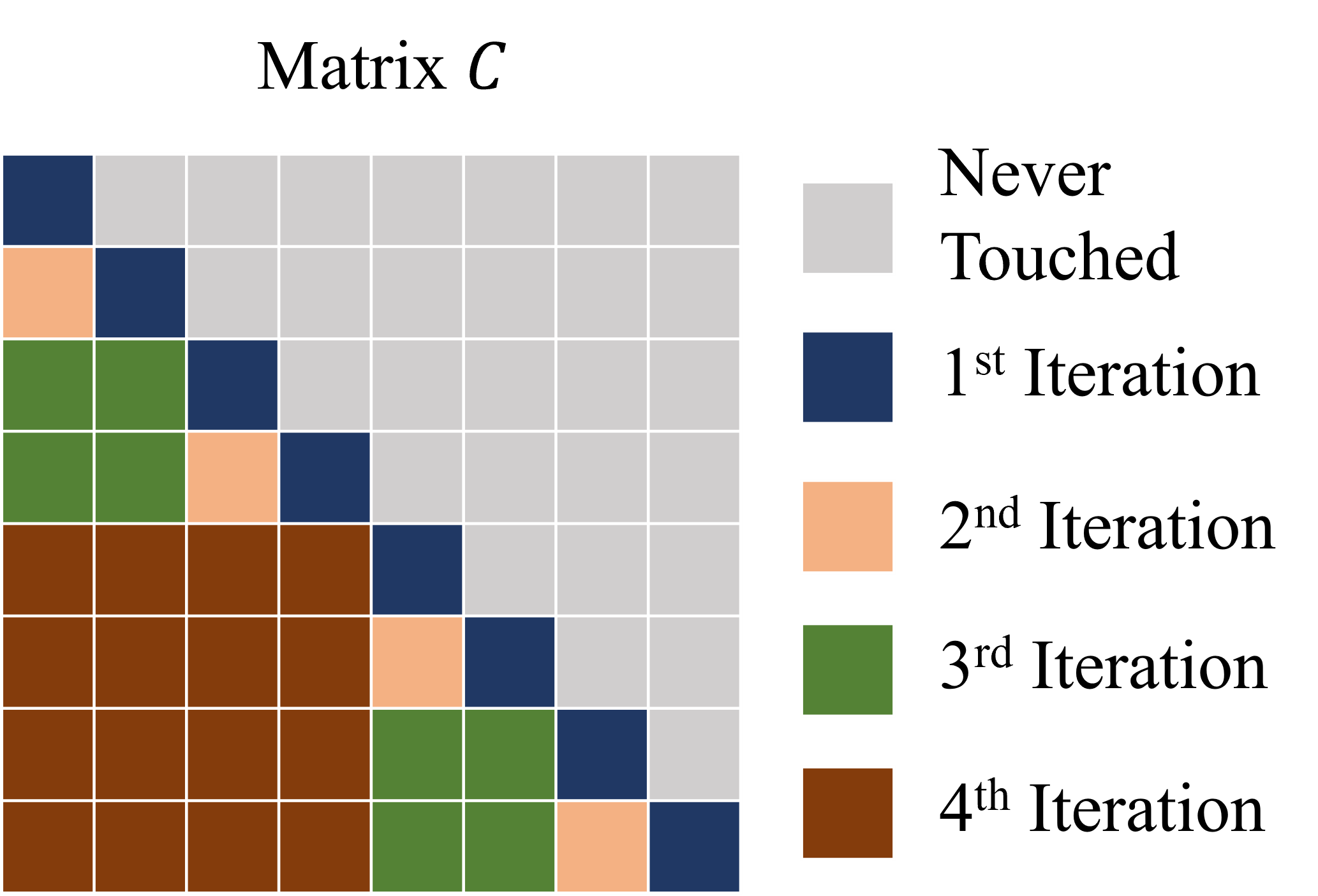}
    \caption{The steps of computing syr2k\label{fig:syr2k_step}}
    
\end{figure}

The original \verb|syr2k| decomposes into two sub-\verb|syr2k| operations and one large GEMM. Recursive decomposition improves hardware utilization by creating larger GEMMs. However, pure recursion reduces parallelism. Therefore, we adapt the recursive idea into a parallel version, ensuring sub-\verb|syr2k| operations run in parallel. Figure~\ref{fig:syr2k_step} illustrates the iterative approach for computing \verb|syr2k|. In the first iteration, diagonal blocks are computed in parallel (using cuBLAS batched GEMM), followed by iterative computation of off-diagonal blocks until the largest GEMM is processed. See Algorithm~\ref{alg:rsyrk} for details.

\begin{algorithm}
\caption{Recursive-like Symmetric Rank-2k Update with blocksize \textbf{nb}}\label{alg:rsyrk}
\begin{Verbatim}[numbers=left,xleftmargin=5mm,commandchars=\\\{\}]
function [C] = \textcolor{blue}{syr2k}(A, B)
  [n,k] = size(A);
  \textcolor{red}{%BatchedGEMM((m,n,k),A,B,offset)}
  \textcolor{red}{%1st iteration}
  C1+=\textcolor{blue}{BatchedGEMM}((nb,nb,k),A,B',nb*(1+lda));
  C1+=\textcolor{blue}{BatchedGEMM}((nb,nb,k),B,A',nb*(1+lda));
  i=1;
  while(n/nb/i/2>=1)
    \textcolor{red}{%2nd to n-th iteration}
    Ci+=\textcolor{blue}{BatchedGEMM}((i*nb,i*nb,k),A+i*nb,
                   B',2*i*nb(1+lda));
    Ci+=\textcolor{blue}{BatchedGEMM}((i*nb,i*nb,k),B+i*nb,
                   A',2*i*nb(1+lda));
    i=i*2;
  end
  C=\textcolor{blue}{concat}(C1,C2,...,Cn);
end
\end{Verbatim}
\end{algorithm}

\begin{figure}
\begin{subfigure}[b]{1.0\columnwidth}
    \includegraphics[width=\textwidth]{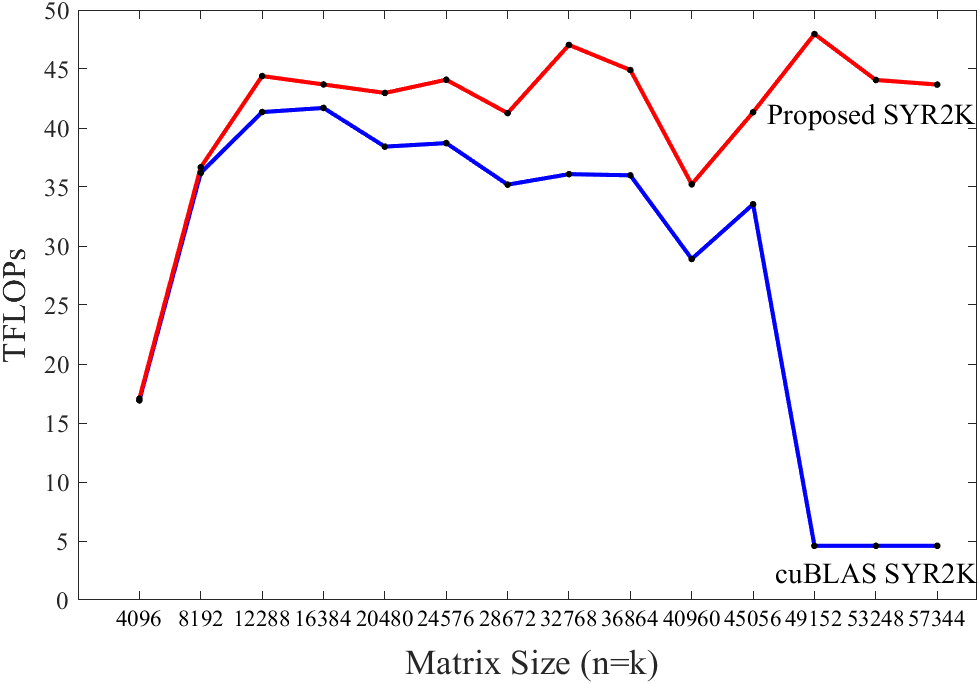}
    \caption{The SYR2K performance comparison between the proposed method and cuBLAS, the input $A$ and $B$ are square with size $n\times n$}
     \label{fig:syr2k_square}
\end{subfigure}

\begin{subfigure}[b]{1.0\columnwidth}
    \includegraphics[width=\textwidth]{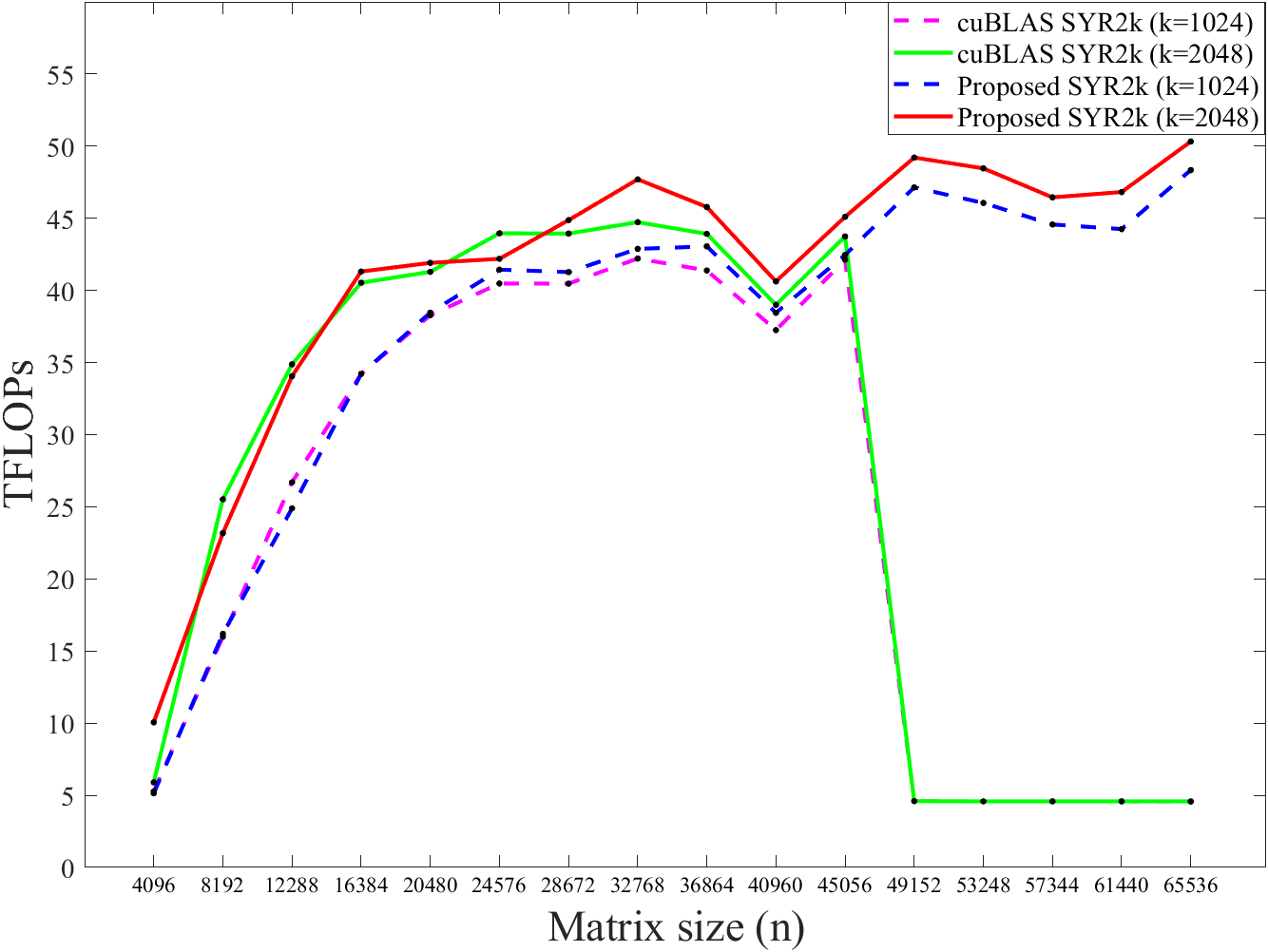}
    \caption{The FP64 precision roofline model of 4 GPU generations}
     \label{fig:syr2k_ts}
\end{subfigure}

\caption{The performance comparison between the proposed SYR2K and cuBLAS SYR2K with different shapes and sizes of input matrices on H100 GPU}
\label{fig:syr2k_comp}
\end{figure}

The speedup compared to cuBLAS \verb|Dsyr2k| routine is shown in Figure~\ref{fig:syr2k_comp}. We compared the square \verb|SYR2K| and tall and skinny \verb|SYR2K| and the experiments reveal that we can outperform cuBLAS on varies sizes and shapes.

% \subsubsection{Tall and Skinny GEMMs Optimization}

% Another optimization is accelerating the tall and skinny GEMMs in the computations between $b$ and $nb$. Although mathematical operations of these computations is much lower than \verb|syr2k|, but the execution rate is quite low (around 1-2 TFLOPs on H100 GPU), which results in 40\% of the total tridiagonalization time cost.

\subsection{Bulge Chasing Kernel}
From our previous analysis and Algorithm~\ref{alg:bulge_chasing}, the design strategy for bulge chasing is straightforward: each kernel handles one sweep.

Within the kernel, multiple threads perform the Householder transformations from both the left and right sides, with the number of threads determined by the bandwidth $b$. Given that bulge chasing is memory-bound, it is crucial to hide the data movement between shared and global memory. We achieve this by maintaining two shared memory blocks: one for online computations and another for data movement. Once the computation block completes its tasks, its role switches to data movement, while the other block takes over computations. This simple pipelining allows for overlapping some data movements.

A critical issue between different sweeps (kernels) is synchronization due to data dependencies between successive sweeps. Specifically, the $i+1$-th sweep can only commence after the third bulge elimination in the $i$-th sweep. Nvidia’s cooperative group tool facilitates synchronization between threadblocks. However, for this to function correctly, the number of launched blocks must not exceed the device limit, or the cooperative group will return an error indicating "too many blocks in cooperative launch cudaLaunchCooperativeKernel."

In our implementation, to maximize parallelism, the bulge chasing process launches a large number of threadblocks, particularly for large matrices. Therefore, using a cooperative group is impractical in this scenario. To resolve the synchronization issue, we manually set up locks between adjacent sweeps to avoid conflicts. The $i$-th sweep shares a lock flag with the $i+1$-th sweep to indicate the current column being processed. After the $i+1$-th sweep eliminates a bulge, it waits until there is no conflict as indicated by the lock. Similarly, the $i+2$-th sweep monitors the lock flag shared with the $i+1$-th sweep.

\begin{figure}
    \centering
    \includegraphics[width=1.0\columnwidth]{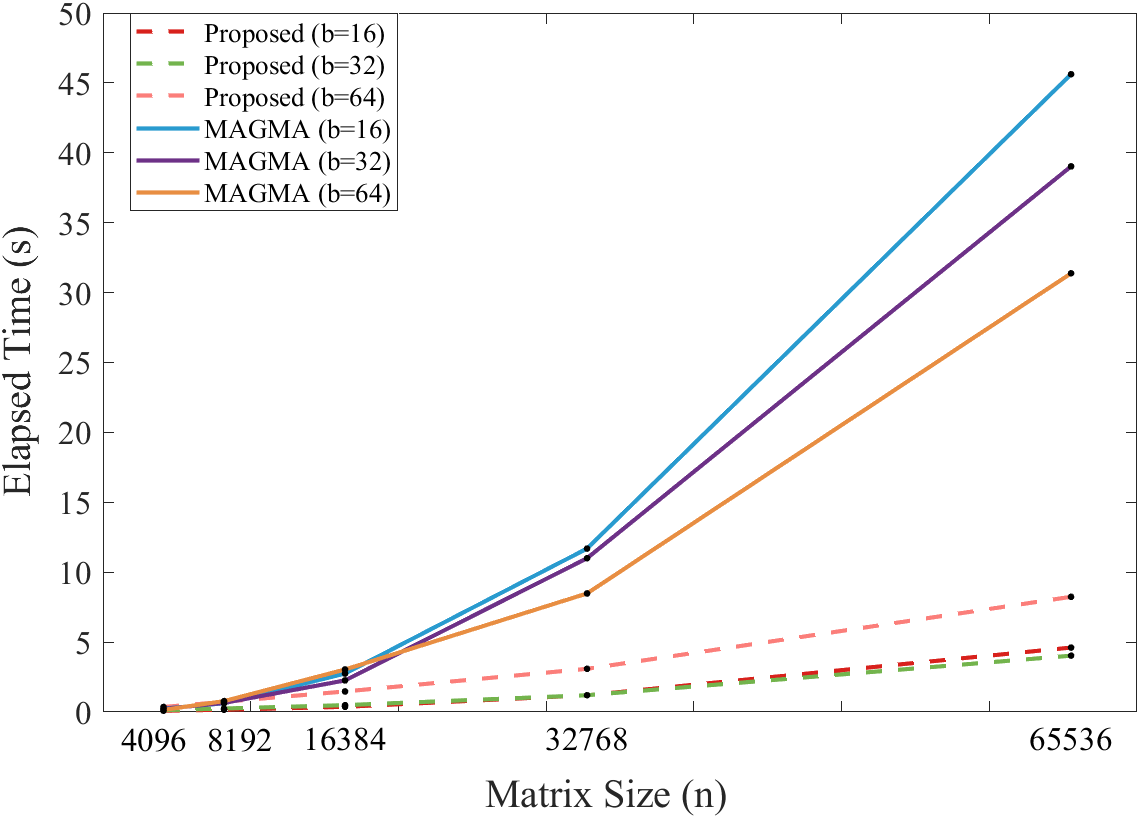}
    \caption{The bulge chasing performance comparison between MAGMA and our proposed implementation with different matrix sizes and bandwidth}
     \label{fig:bc_perf_comp}
\end{figure}

Our proposed bulge chasing implementation and optimization take full advantage of hardware accelerators, as shown by the performance comparison with MAGMA in Figure~\ref{fig:bc_perf_comp}. The proposed implementation achieves 8.0x speedup compared to the CPU-based version.

\subsection{Tuning}
Like conventional SBR, tuning parameters such as $b$ and $nb$ is essential to achieve optimal performance on different hardware. The bandwidth $b$ is crucial for balancing bulge chasing and trailing matrix updates. In our DBR, $nb$ must also be carefully selected to ensure peak performance during trailing matrix updates. For instance, on an A100 GPU, setting $nb=512$ is sufficient for high-performance \verb|syr2k| operations, and increasing $nb$ further would result in slower panel updates. Conversely, on an H100 GPU, \verb|syr2k| operations require $nb \geq 2048$, so we set $nb=2048$. In general, the block size $nb=k$, where $k$ in \verb|syr2k| operation yields the best performance for large matrices.

Although we have adapted panel updates using a recursive formulation, the bandwidth $b$ does not significantly impact DBR performance, as many tall and skinny GEMMs are converted into relatively square GEMMs. However, GEMMs with one small dimension still exist and are non-negligible. Additionally, bulge chasing performance is closely related to bandwidth $b$, with smaller $b$ being preferred. Thus, it is important to find a balance between DBR and bulge chasing through tuning. Fortunately, DBR is not highly sensitive to tuning, and even suboptimal settings can deliver satisfactory performance. See Table~\ref{tbl:tuning}) for a performance reference.

\begin{table}[]
   
\centering
\begin{tabular}{ |c|c|c|c|c|c|c||c|c| } 
\hline
\diagbox{$b$}{$nb$} &  128 & 256 & 512 & 1024 & 2048 & 4096 & BC\\
 \hline
 
16 & 31.9 & 26.3 & 23.8 & 23.4 & 23.7 & 25.4& 4.6\\
\hline
32 & 24.0 & 18.8 & 16.4 & 15.6 & 15.5 & 16.2 & 4.0\\
\hline
64 & 20.2& 15.0& 12.6& 11.7& 11.4 & 11.6 & 8.2\\
\hline
\end{tabular}
\caption{The elapsed time in seconds of DBR and bulge chasing (BC, the last column) on H100 GPU with matrix size $65536\times 65536$ }
\label{tbl:tuning}
\end{table}

\subsection{Summary}
In this section, we depict our implementations and optimizations of the tridiagonalization process, including DBR, BLAS3 operations, and bulge chasing. We've observed significant acceleration compared to MAGMA on the H100 GPU.

The final step of the EVD solver is computing the eigenvalues using iterative methods such as QR algorithm. As previously discussed, the iterative method is not a bottleneck on GPUs. For example, cuSOLVER's well-optimized divide and conquer only takes about 3\% of the time in the \verb|Dsyevd| routine. We acknowledge that outperforming cuSOLVER's implementation in this regard is unlikely. However, due to the tremendous speedup in tridiagonalization, even a sub-optimal divide and conquer implementation can still result in substantial acceleration of the full eigenvalue decomposition.

Another observation is that our proposed methods are also beneficial on emerging GPU architectures such as the RTX 4090 series, which have limited FP64 precision computing capacity (only $\frac{1}{64}$ of FP32 peak performance). Our methods can take advantage of new technologies that leverage INT8 Tensor Cores to perform FP64 GEMMs~\cite{TCDgemm}. Performance results will be shown in the experiments section.

\section{Experimental Evaluation}

We conducted experiments on a system running a 5.4.0-99-generic Linux operating system with NVIDIA H100-SMX, A100-PCIe, and RTX 4090 GPUs. The CUDA version used is 12.3, which includes a C++ compiler and the cuBLAS and cuSOLVER libraries. This section will primarily showcase the performance of the proposed tridiagonalization and the entire eigenvalue decomposition on various GPU architectures.

\begin{figure}
\begin{subfigure}[b]{1.0\columnwidth}
    \includegraphics[width=\textwidth]{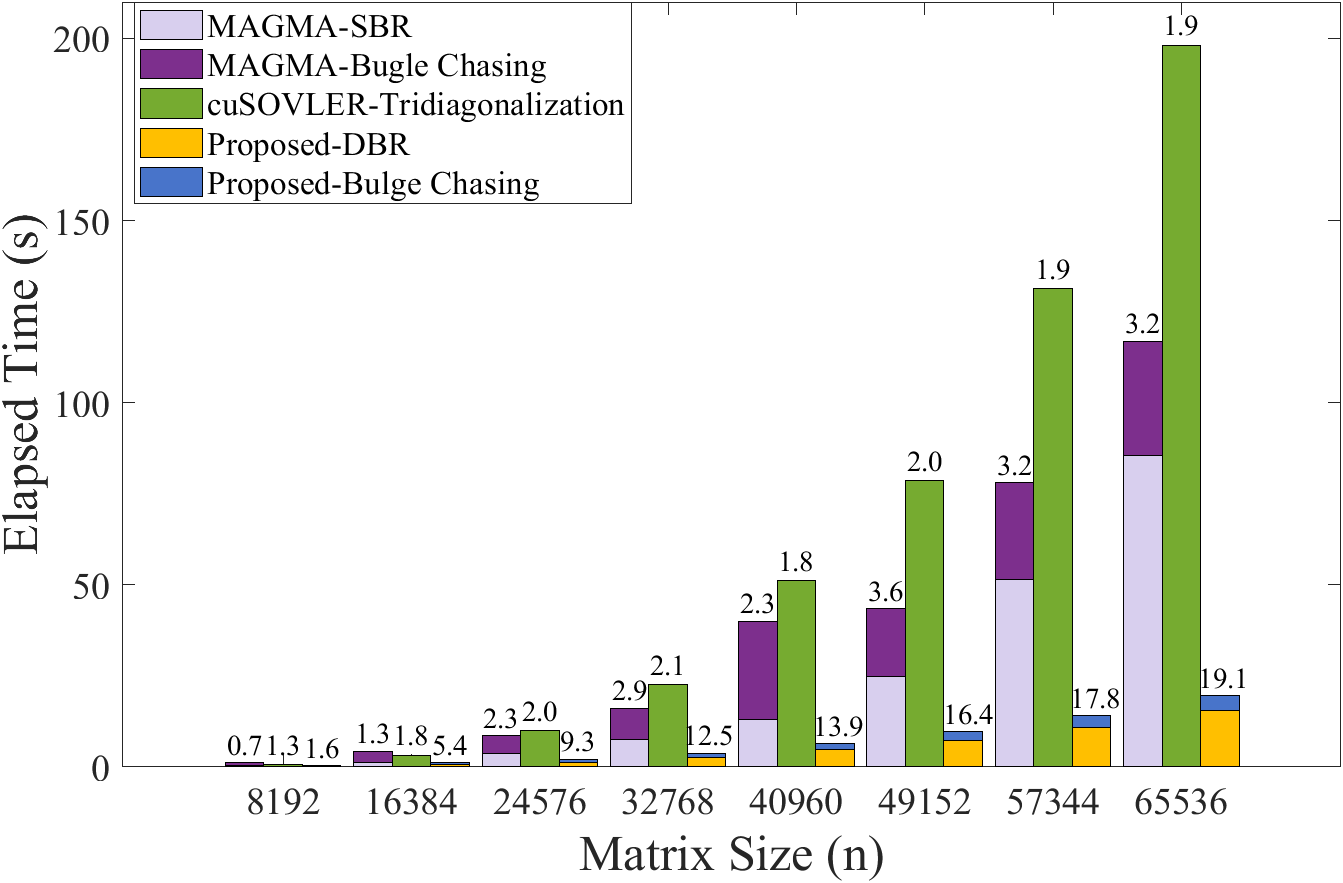}
    \caption{The tridiagonalization performance on H100 GPU}
     \label{fig:tr_perf_h100}
\end{subfigure}

\begin{subfigure}[b]{1.0\columnwidth}
    \includegraphics[width=\textwidth]{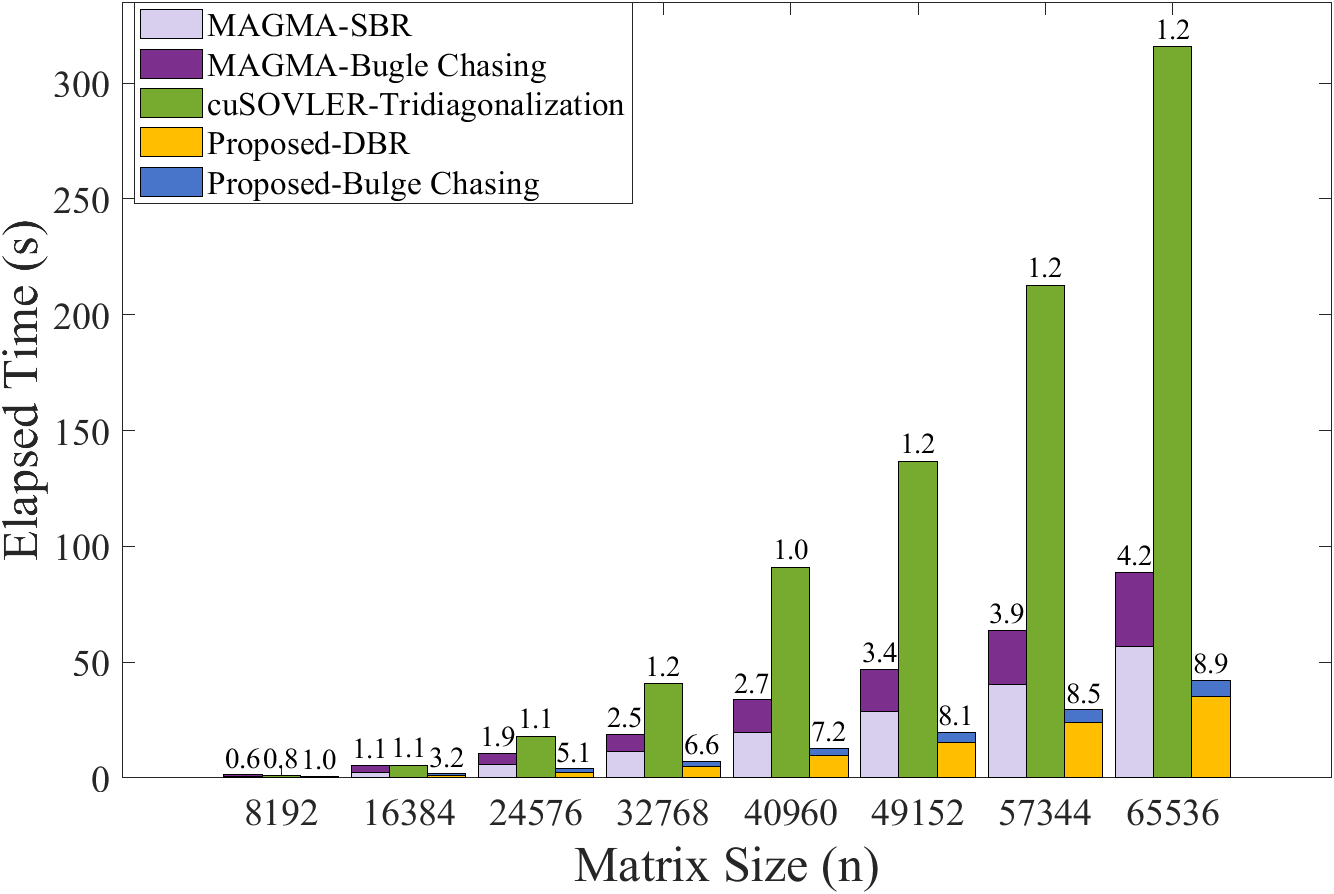}
    \caption{The tridiagonalization performance on A100 GPU}
     \label{fig:tr_perf_a100}
\end{subfigure}

\begin{subfigure}[b]{1.0\columnwidth}
    \includegraphics[width=\textwidth]{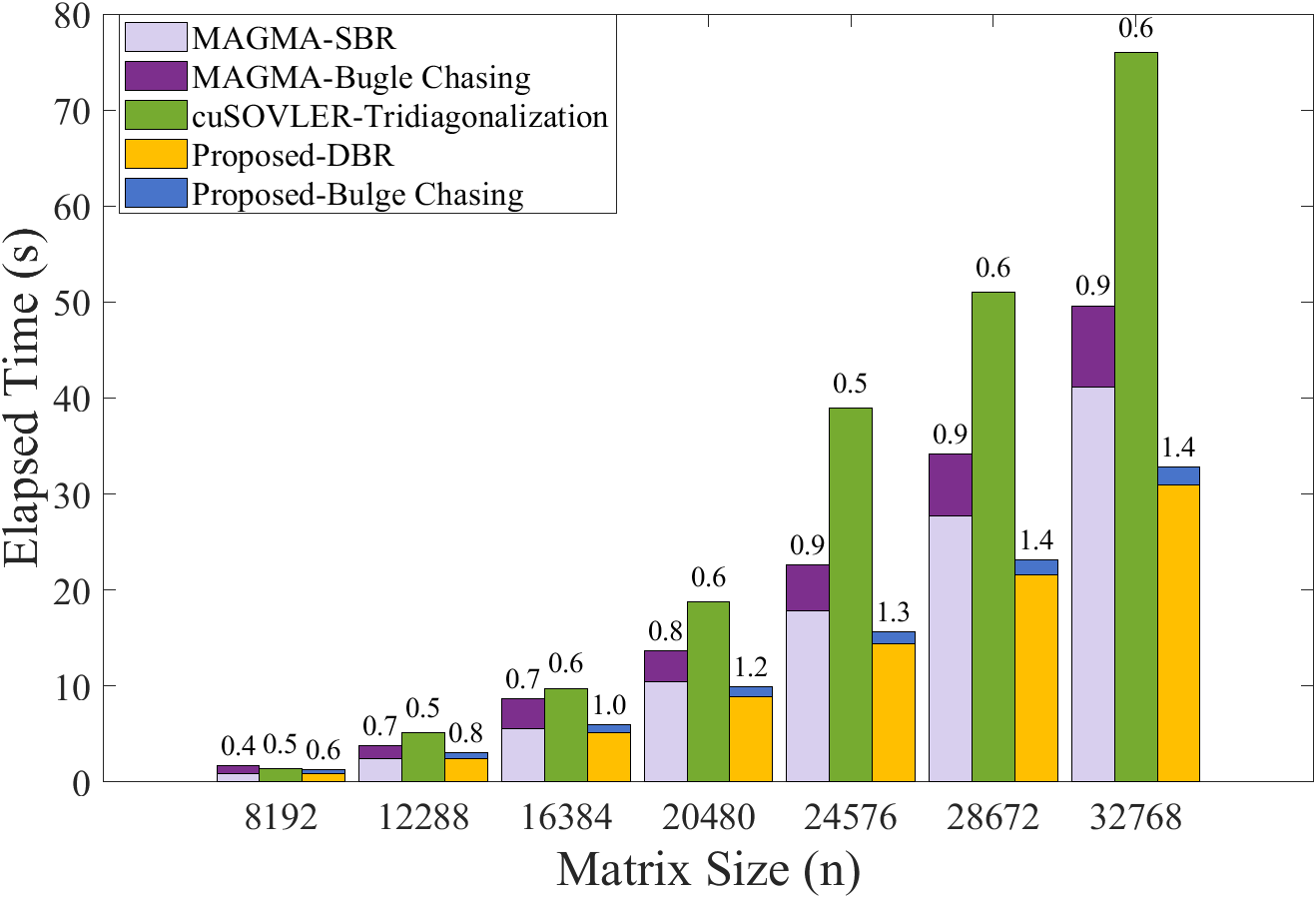}
    \caption{The tridiagonalization performance on RTX 4090 GPU}
     \label{fig:tr_perf_4090}
\end{subfigure}

\caption{Tridiagonalization performance comparison among cuSOLVER, MAGMA, and our proposed method for different matrix sizes. The numbers on top of the bars denote TFLOPs, with the peak FP64 performance of the H100, A100, and RTX 4090 GPUs being 67 TFLOPs, 19.5 TFLOPs, and 1.25 TFLOPs, respectively}
\label{fig:tr_comp}
\end{figure}

\begin{figure}
    \centering
    \includegraphics[width=1.0\columnwidth]{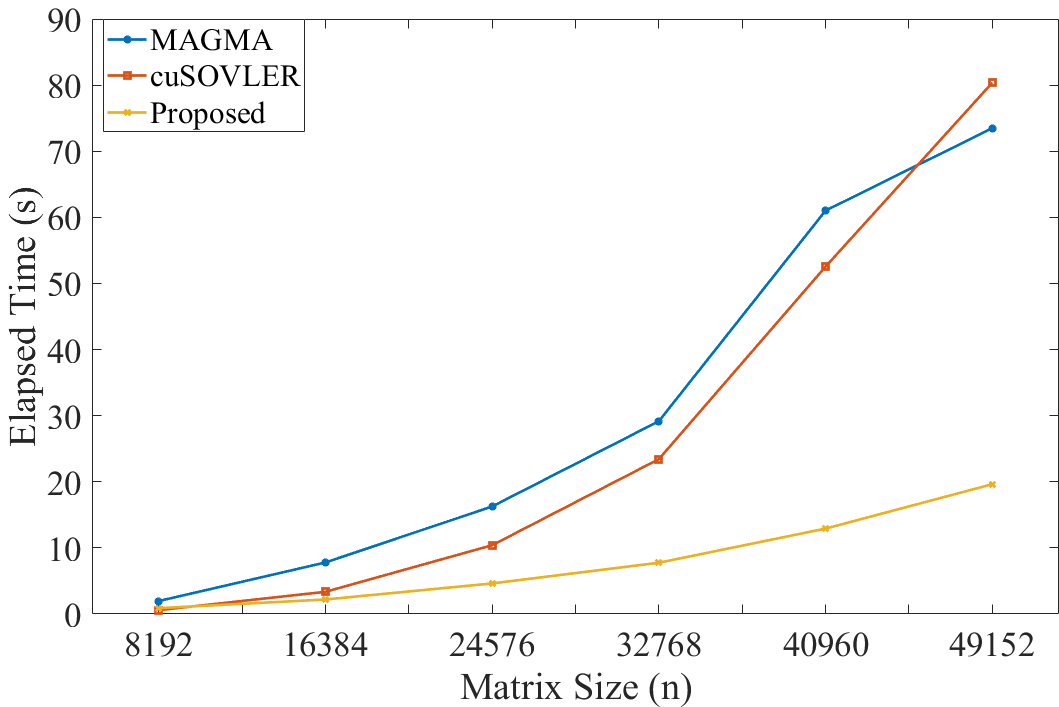}
    \caption{The EVD solver comparison between MAGMA, cuSOVLER, and our proposed EVD solver on H100 GPU}
     \label{fig:evd_per_h100}
\end{figure}

Evaluating our algorithmic design requires comparing performance metrics against cuSOLVER and MAGMA across various matrix sizes and GPU architectures. In previous sections, we validated our sub-modules on the H100 GPU. However, this alone is insufficient to fully demonstrate the proposed algorithm's capabilities. Here, we extend our evaluation to include the A100 GPU, which exhibits lower ratio between computing capacity and memory bandwidth. Additionally, we demonstrate that our algorithm can leverage INT8 Tensor Cores~\cite{TCDgemm} to produce FP64 precision results on consumer-grade GPUs such as the RTX 4090, which offers limited support for FP64 precision.

Figure~\ref{fig:tr_comp} gives the performance comparison of tridiagonalization among cuSOLVER, MAGMA, and our proposed method on the H100, A100, and RTX 4090 GPUs. The experimental results reveal that our method outperforms cuSOLVER (the SOTA GPU-only implementation) and MAGMA (the SOTA two-stage tridiagonalization on a CPU-GPU hybrid architecture) across a range of matrix sizes and GPU architectures. This performance enhancement is attributed to our method's higher data intensity, which leads to better utilization of hardware accelerators.

In terms of TFLOPs, our proposed method achieves up to 29\% and 46\% of peak FP64 performance on the H100 and A100 GPUs, respectively. In contrast, cuSOLVER only reaches 2.8\% and 6.2\% of peak performance on these GPUs. While MAGMA, benefiting from its two-stage tridiagonalization algorithm, is slightly faster than cuSOLVER for large matrices, it remains inefficient on hardware accelerators, utilizing only approximately 4.7\% and 21.5\% of peak performance. On the RTX 4090 GPU, leveraging the power of INT8 Tensor Cores, our method surpasses the FP64 performance limit (1.25 TFLOPs) imposed by the hardware, a feat unachievable by cuSOLVER and MAGMA, which cannot exploit Tensor Cores.

We also evaluate the end-to-end eigenvalue decomposition (EVD) using the divide-and-conquer method. The implementation of iterative methods involves more technical than algorithmic innovations, making it challenging to outperform cuSOLVER, as Nvidia researchers possess deep expertise in optimizing on Nvidia's hardware accelerators. Nonetheless, our significantly faster tridiagonalization step contributes substantially to end-to-end acceleration. Despite our suboptimal divide-and-conquer implementation, we achieve approximately a 4.1x speedup on the EVD solver when only eigenvalues are required (cuSOLVER cannot handle larger matrices due to memory limitations), as detailed in Figure~\ref{fig:evd_per_h100}. If we have access to cuSOLVER's implementation, we anticipate even greater speedups. MAGMA's performance is hindered by its iterative methods, as it employs the QR algorithm on the CPU, which has limited parallelism. Thus, MAGMA only outperforms cuSOLVER for very large matrices, whereas our proposed method surpasses cuSOLVER at smaller sizes, despite our less efficient divide-and-conquer algorithm.

\section{Related Work}

\subsection{The Development of Modern Hardware Accelerators}

The boost of hardware accelerators' developments began with Nvidia's release of the P100 GPU in 2016, which supported half-precision computations~\cite{P100}. Driven by the increasing demand for training larger neural networks, the Tesla V100 GPU~\cite{v100Bech} introduced Tensor Cores~\cite{V100}, enabling exceptionally fast fused GEMMs. High precision computations have similarly benefited from these innovations. On the A100~\cite{A100} and H100 GPUs~\cite{H100}, FP64 GEMMs can also utilize Tensor Cores, achieving performance levels equivalent to FP32 computations. Specifically, for scientific computing requiring high precision, AMD's recently introduced MI300x~\cite{MI300} offers the highest FP64 computing capacity, with a peak performance almost three times that of the H100 GPU, reaching 163.4 TFLOPs.

\subsection{Eigenvalue Decomposition on Modern Computer Architectures}

\subsubsection{Tridiagonalization}
% The most widely used and well-known algorithm is the QR algorithm~\cite{QRAlgorithm}, which repeatedly calls the QR factorization and GEMM and finally converges to a diagonal matrix that contains eigenvalues on its diagonal. 

% It's well know that it's impossible to express  roots of general high order polynomial in radicals, it's impossible to directly factorize EVD. Therefore, the iterative method such as QR algorithm~\cite{QRAlgorithm}, divide and conquer~\cite{gu1995divide}, and Jacobi iterations~\cite{golub2000eigenvalue}.
On current computer architectures, tridiagonalization is typically regarded as a 'preconditioner' to reduce the computational time of iterative methods. Tridiagonalization is commonly performed using Householder transformations~\cite{MatrixComputation}. To enhance execution efficiency on modern high-performance architectures, the WY representation technique~\cite{WYRepresentation, WYStorage} is often applied during the transformation process. To further improve data locality, two-stage tridiagonalization~\cite{2stage1} is frequently employed for larger matrices. This approach first reduces the matrix to a band form (SBR) and then reduces the band form to a tridiagonal matrix using bulge chasing. This method has been demonstrated to be highly efficient on multi-core architectures~\cite{2stage2, 2stage3, 2stagesvd}.

\subsubsection{Iterative Methods}
It is well known that expressing the roots of general high-order polynomials in radicals is impossible, meaning that a direct EVD solver does not exist. Consequently, eigenvalues must be computed using iterative methods such as the QR algorithm~\cite{QRAlgorithm}, divide and conquer~\cite{gu1995divide}, and Jacobi iterations~\cite{golub2000eigenvalue}. Among these, the divide and conquer method is particularly popular due to its superior parallelism and efficiency in computing eigenvectors. For computing eigenvalues alone, the QR algorithm is often the best choice. These methods are implemented in most linear algebra packages, including LAPACK~\cite{LAPACK}, Eigen~\cite{EigenLibrary}, MAGMA~\cite{MAGMA}, and cuSOLVER.

A flexible method is bisection~\cite{Bisection}, which aims to find a subset of eigenvalues, such as the largest or smallest 100, or all eigenvalues within an interval $[a,b]$. In 2004, the MRRR (Multiple Relatively Robust Representations) method~\cite{MRRR} was introduced. It aims to compute accurately orthogonal eigenvectors without requiring the expensive reorthogonalization process, which has a worst-case complexity of $O(n^3)$.

Another class of iterative methods is based on polar decomposition that links EVD and SVD. The QDWH-eig (QR-based dynamically weighted Halley eigenvalue decomposition)\cite{QDWH} method uses QR factorization to compute the polar decomposition, followed by factorizing the derived orthonormal matrix using an iterative subspace method. In 2016, a GPU implementation\cite{QDWHGPU} of QDWH-eig and QDWH-SVD was proposed. Another method for computing polar decomposition, called scaled Newton~\cite{ScaledNewton}, involves fewer mathematical operations than QDWH. However, it relies heavily on the backward stable inversion of a matrix.

Recently, there has been growing interest in randomized linear algebra~\cite{Randomized}, particularly randomized subspace iteration for computing low-rank approximate eigenvalue or singular value decomposition. Two notable algorithms in this field are randomized subspace iteration~\cite{RandomSubspace} and randomized block Lanczos~\cite{BlockLanczos}. These algorithms have proven efficient in real-world applications, especially on modern high-performance architectures~\cite{xsvm, tensorsvm}. However, these methods typically involve multiplying a randomly generated matrix, which limits their applicability to scenarios where accuracy is not critically sensitive.

\section{Conclusion and Future Work}
In this paper, we discuss the features of the emerging hardware accelerators, that the gap between memory bandwidth and computing capacity of hardware accelerators has widened compared to older architectures. This disparity presents challenges in designing algorithms that fully utilize computing capacity, as the conventional algorithm might become memory-bound. Consequently, the performance of EVD solvers on emerging GPU architectures has been suboptimal, achieving only 1.8 TFLOPs out of a possible 67 TFLOPs on the H100 GPU.

To harness the potential of emerging hardware accelerators, we use EVD solver as an example to show the algorithmic design of how to leverage the computing capacity and parallelism of new hardware. In short, we propose a new band reduction algorithm named detached band reduction, which provides much higher performance on \verb|syr2k|, and we use a recursive-like algorithm for \verb|syr2k| to further improve the performance. For the bulge chasing process in tridiagonalization, we provide a GPU-based implementation which has a better utilization of parallelism on new  Experimental evaluations revealed that our proposed method achieves up to a 10.1x speedup on the H100 GPU and can benefit from INT8 Tensor Cores on consumer-level GPUs. However, the profiling in this paper is still not sufficient that we only compare the TFLOPs, providing more details such as memory footprint will be considered from the hardware aspect in the future.

However, the EVD solver is only a small portion of numerical linear algebra. Other problems, such as SVD solvers and one-sided matrix factorizations, might also benefit from the proposed ideas. Therefore, we will try expand our approach to other numerical linear algebra and matrix computation problems including LU, Cholesky and QR factorization. Using INT8 Tensor Cores to accelerate the above problems on hardware which has limited FP64 computing capacity is also appealing. Additionally, while this paper focuses on the implementations using single hardware accelerator, scaling these problems on emerging clusters is another interesting topic for future exploration. Also, we're targeting on high precision EVD solver in this paper, exploring the possibilities of high performance EVD in low precision using Tensor Cores is another topic, because the deep learning communities might prefer half precision and stable EVD solvers.

\bibliography{ref}

\appendix

\end{document}